\documentclass[a4paper,fleqn,usenatbib]{mnras} 
\usepackage[T1]{fontenc}
\usepackage{ae,aecompl}

\usepackage{graphicx}	
\usepackage{amsmath}	
\usepackage{amssymb}	
\usepackage{wasysym}
\usepackage{breakurl}
\usepackage{color}
\usepackage{pdflscape}

\usepackage[colorinlistoftodos]{todonotes}

\title[The RoboPol four-channel polarimeter]
  {RoboPol: A four-channel optical imaging polarimeter}
\author[A.\,N.~Ramaprakash et al.]
  {A.\,N.~Ramaprakash,$^1$\thanks{Email: anr@iucaa.in}
  C.\,V.~Rajarshi,$^1$ 
  H.\,K.~Das,$^1$ 
  P.~Khodade,$^1$ 
  D.~Modi,$^1$
  \newauthor
  G.~Panopoulou,$^2$\thanks{Email: panopg@caltech.edu}
  S. Maharana,$^1$
  D. Blinov,$^{3, 4, 5}$
  E.~Angelakis,$^6$
  C. Casadio$^{6}$,
  \newauthor
  L.~Fuhrmann$^{6}$,
  T.~Hovatta$^{7, 8}$,
  S. Kiehlmann$^2$,
  O. G. King,
  N.~Kylafis$^{3, 4}$,
  \newauthor
  A.~Kougentakis$^3$,
  A.~Kus$^{9}$,
  A. Mahabal$^2$,
  A. Marecki$^9$, 
  I.~Myserlis,$^6$
  \newauthor
  G.~Paterakis,$^3$
  E.~Paleologou$^{3}$,
  I. Liodakis$^{10}$,
  I.~Papadakis$^{3, 4}$,
  I.~Papamastorakis$^{3, 4}$,
  \newauthor
  V.~Pavlidou$^{3, 4}$,
  E.~Pazderski$^{9}$,
  T.\,J.~Pearson$^{2}$,
  A.\,C.\,S.~Readhead$^{2}$,
  \newauthor
  P.~Reig$^{3, 4}$,
  A. S\l{}owikowska$^{9}$,
  K.~Tassis,$^{3, 4}$
  and J.\,A.~Zensus$^6$\\
  $^1$Inter-University Centre for Astronomy \& Astrophysics, Post bag 4, Ganeshkhind, Pune - 411007, India \\
  $^2$Cahill Center for Astronomy and Astrophysics, California Institute of Technology, 1200 E California Blvd, MC 249-17, \\Pasadena CA, 91125, USA\\
  $^{3}$Foundation for Research and Technology - Hellas, IA \& IESL, Voutes, 71110 Heraklion, Greece \\
  $^4$ Department of Physics and Institute of Theoretical \& Computational Physics, University of Crete, PO Box 2208, \\GR-710 03, Heraklion, Crete, Greece \\
  $^5$ Astronomical Institute, St. Petersburg State University, Universitetsky pr. 28, Petrodvoretz, 198504 St. Petersburg, Russia\\
  $^6$Max-Planck-Institut f\"{u}r Radioastronomie, Auf dem H\"{u}gel 69, 53121 Bonn, Germany\\
  $^{7}$Finnish Centre for Astronomy with ESO (FINCA), University of Turku, FI-20014, Turku, Finland\\
  $^{8}$Aalto University Mets\"ahovi Radio Observatory, Mets\"ahovintie 114, 02540 Kylm\"al\"a,Finland\\
 $^{9}$Toru\'{n} Centre for Astronomy, Nicolaus Copernicus University, Faculty of Physics,
Astronomy and Informatics, \\ Grudziadzka 5, 87-100 Toru\'{n}, Poland \\
 $^{10}$KIPAC, Stanford University, 452 Lomita Mall, Stanford, CA 94305, USA \\
}

\begin{document}

\date{Accepted XXX. Received YYY; in original form ZZZ}

\pagerange{\pageref{firstpage}--\pageref{lastpage}} \pubyear{2019}

\maketitle

\label{firstpage}

\begin{abstract}
We present the design and performance of RoboPol, a four-channel optical polarimeter operating at the Skinakas Observatory in Crete, Greece. RoboPol is capable of measuring both relative linear Stokes parameters $q$ and $u$ (and the total intensity I) in one sky exposure. Though primarily used to measure the polarization of point sources in the R-band, the instrument features additional filters (B, V and I), enabling multi-wavelength imaging polarimetry over a large field of view ($13.6\arcmin \times13.6\arcmin$). We demonstrate the accuracy and stability of the instrument throughout its five years of operation. Best performance is achieved within the central region of the field of view and in the R band. For such measurements the systematic uncertainty is below 0.1\% in fractional linear polarization, $p$ (0.05\% maximum likelihood). Throughout all observing seasons the instrumental polarization varies within 0.1\% in $p$ and within $\sim$1$^\circ$ in polarization angle.
\end{abstract}


\section{Introduction} \label{sec:introduction}

Modern polarimeter design is driven by diverse science goals. Examples include the aim to detect the extremely low polarization signature of planets near bright stars \citep{Hough2006,Wiktorowicz2015}, or the wavelength-dependence of time-varying polarization through synchronous multi-band imaging \citep{Piirola2014}.

The most commonly used design for imaging polarimetric instruments is the dual-beam polarimeter \citep[e.g.][]{1967PASP...79..136A}, which at its heart combines a modulator/retarder (e.g. rotating half-wave plate) with a beam analyser (e.g. birefringent prism). 
Compared to its predecessor, the single-beam polarimeter, this type of instrument offers the advantage of cancellation of multiplicative noises that affect the two beams (such as variations in atmospheric opacity between exposures) \citep{scarrott1983}.
This design only allows for measurement of one polarized Stokes parameter ($Q$ or $U$ for linear polarimetry) at a time. In order to obtain the fractional linear polarization, $p$, and polarization position angle, $\chi$, (and also to break the mirror degeneracy of the latter), at least two exposures at different orientations of the half-wave plate (usually at 0$^\circ$ and 22.5$^\circ$) are necessary. In practice, to correct for instrumental effects such as differential response of the polarimeter to different polarization states of incoming light, two additional exposures are taken at 45$^\circ$ and 67.5$^\circ$ \citep[e.g.][]{Magalhaes1996,Ramaprakash}. So a typical polarization measurement with a dual beam polarimeter consists of 4 exposures at different half-wave plate positions. Dual beam polarimetry is susceptible to target variations between exposures, incorrect alignment of the half-wave plate and can have large CCD readout time overheads. 

These constraints are bypassed in the quadruple-beam (or four-channel) polarimeter design. First proposed by \cite{Geyer1996}, the four-channel polarimeter uses a pair of birefringent prisms (in this case Wollaston prisms, WP) as the beam analyzer to achieve simultaneous measurement of both Stokes $Q$ and $U$. Four beams polarized at 0$^\circ$, 45$^\circ$, 90$^\circ$ and 135$^\circ$ emerge out of this prism pair. The measurement of the relative intensities of the first two beams provides the Stokes parameter $Q$ (0$^\circ$, 90$^\circ$) and relative intensities of the other two beams provides $U$. The prisms are placed in such a way that the telescope beam is shared approximately equally between the two, and all polarization states are imaged simultaneously at different positions on the detector. This basic principle has been implemented in a number of existing instruments \citep[e.g.][]{pernechele2003,fujita2009,helhel2015,devogele2017}.

The efficiency of a four-channel instrument at first glance seems inferior to that of the dual-channel design: the light of the source is split into four rays, compared to two in the dual-beam design. However, in reality the four channel design does not lead to any loss of performance, due to the fact that the uncertainty on a Stokes parameter measurement depends on the noise of the total intensity (the sum of the two beams) used to obtain the Stokes parameter, and not that of a single beam. Thus, a four-channel instrument can achieve the same accuracy, in terms of photon noise, as a (perfect) dual-beam instrument in only twice the time. However, the control of systematics in dual-beam polarimetry, which requires a minimum of 4 exposures, results in the same amount of exposure time as in the four-channel case. With the four-channel setup, any inaccuracies due to the positioning of a rotating retarder, or other systematics due to instrument changes between exposures, are avoided at no cost in terms of timing.

RoboPol is a four channel polarimeter, capable of measuring the linear Stokes parameters in one exposure. A collimated telescope beam is shared equally by two quartz Wollaston prisms, each with its own half-wave plate in front. Four beams of differing polarization states are output from this system, and are imaged on a CCD detector. Relative photometry of the four beam images provides the linear Stokes parameters. RoboPol is mounted on the 1.3 m telescope of the Skinakas observatory in Crete, Greece. The instrument was custom designed and built to conduct a comprehensive long term blazar polarimetric monitoring campaign.

Operating successfully since 2013, data collected by RoboPol have contributed in multiple publications in the field of blazars \citep{Pavlidou2014,Blinov2015,Blinov2016a,Blinov2016b,Hovatta2016,Angelakis2016,Liodakis2017,Raiteri2017,Kiehlmann2017,Uemura2017,Blinov2018}, galactic binaries and white dwarfs \citep{Reig2014,Zejmo2017,Reig2017,reig2018,slowikowska2018}, gamma-ray bursts \citep{King2014}, and the interstellar medium \citep{Panopoulou2015,Panopoulou2016,Skalidis2018}. 

We present the instrumental design in section \ref{sec:instrument}. Section \ref{sec:commissioning} describes the commissioning phase. We demonstrate the accuracy and long-term stability of the instrument in section \ref{sec:performance}. We summarize in section \ref{sec:conclusion}.

\section{The Instrument} \label{sec:instrument}

\begin{figure*}
\centering
\includegraphics[scale = 0.7]{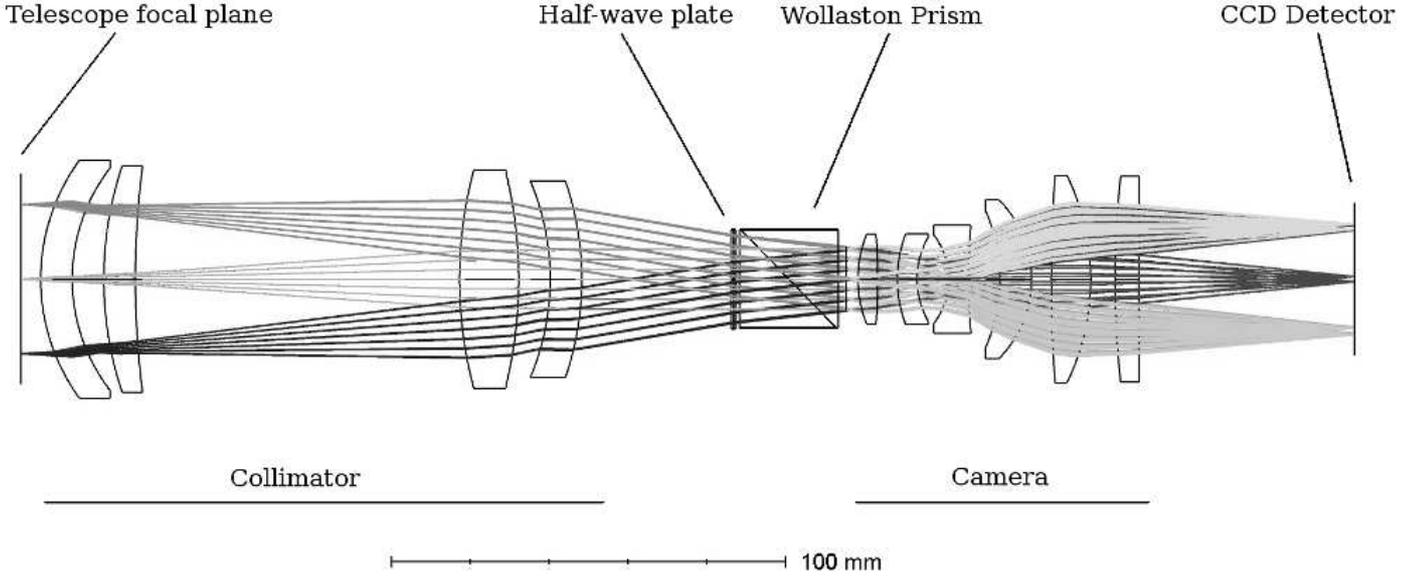}
\caption{Optical design of RoboPol (side view). The optical train from left to right: telescope focal plane, collimator lenses, half-wave plates \& Wollaston Prisms system, camera lenses, CCD. Light rays originating at different locations on the FOV are shown as they propagate through the system (differently shaded gray lines). }
\label{fig:optical_design}
\end{figure*}

\subsection{Design Considerations}

The main scientific driver for building RoboPol was to explore the nature of the coherent rotation of the polarization angle of blazar optical emission, first observed three decades ago \citep[e.g.][]{Kikuchi1988}. Though different mechanisms have been proposed to explain this behavior \citep[e.g.][]{Choudhuri1985,Bjornsson1982,Sillanpaa1993,abdo2010}, observational evidence has lagged behind. This was our main motivation for initiating the RoboPol blazar monitoring campaign. The project goal was to observe the linear optical polarization of a large, statistically unbiased sample of blazars with high cadence for a duration of three years \citep{Pavlidou2014}.

The RoboPol polarimeter has been developed with two main priorities in mind: efficiency and accuracy. The large sample of targets ($\sim$ 100) to be covered with a cadence of several days calls for high observing efficiency (with minimal overheads such as pointing and slewing). The optical emission from blazars is typically linearly polarized at a level of a few percent \citep{Pavlidou2014}. For the limiting case of a source with fractional linear polarization, $p$, of 1\% and R = 18 mag the instrument should be capable of producing a 3-$\sigma$ detection within 30 minutes of exposure time. Systematic uncertainties should therefore be below $\sigma_p$ = 0.3\%.

According to these considerations, a four-channel design was selected for RoboPol as it meets the needs of the scientific program: high measurement accuracy with minimal overheads. In order to minimize the amount of time spent on a specific target, monitoring is performed in a single band (Johnson-Cousins R). 
 
The instrument is also capable of performing complementary science in the B, V and I bands (through the use of a filter wheel). The instrument features a large field of view (FOV) ($13.6\arcmin \times 13.6\arcmin$). This provides the benefit of performing relative photometry for the central target source (to obtain Stokes I). This also enables rapid polarimetric mapping of point sources over large areas on the sky.

\subsection{Telescope and CCD camera} \label{subsec:tel_cam}

The polarimeter has been designed for (and is mounted on) the 1.3 m Ritchey-Chr\'etien telescope at the Skinakas observatory in Crete. 
The telescope\footnote{\url{http://skinakas.physics.uoc.gr/en/}} has the following specifications: 129 cm \diameter{} primary f/2.88, 45 cm \diameter{} secondary f/4.39, telescope f/7.64 and an equatorial mount. To satisfy the spatial needs of the instrument, the telescope's focus was repositioned, by altering the distance between the primary and secondary mirror by 15.281 mm. This resulted in a shift of the focus position of 120 mm towards the primary, setting the telescope to an f-number of f/7.488.

RoboPol is mounted on the direct port of the standard Guidance and Acquisition Module (GAM) at the Cassegrain focal station of the telescope. Along with RoboPol, the telescope is equipped with other instruments, including an imaging camera, an infrared camera, and a spectrograph. The telescope beam can be diverted to one of the side port instruments by a fold mirror on a linear stage. The mirror is stowed out of the way when RoboPol is in use. We avoided using the side port for RoboPol to eliminate instrument polarization that would be introduced by the science fold mirror.

The detector used for RoboPol, provided by the Skinakas Observatory, is an Andor DW436 CCD camera with 2048$\times$2048 pixels of size 13.5 $\mu$m. By use of a Peltier device, the camera can be cooled to $-70^\circ$C ensuring negligible dark current ($6\times10^{-4}$ $e^{-}$/sec/pix). The gain and the readout noise at the 2 $\mu$sec/pix readout speed that we use are 2.687 $e^{-}$/ADU and $8.14\pm0.12$ $e^{-}$ respectively. The median seeing at Skinakas is 1$\arcsec$.
At the camera the focal ratio of f/5 gives a mean pixel angular size of $\sim$0.4\arcsec, and a FOV of 13.6\arcmin.

The defocusing of the secondary leads to image quality degradation at the telescope focal plane. However, this is adequately accounted for and compensated by the instrument's optics so that the final delivered image quality on the CCD meets the design requirements.

\begin{figure}
\centering
\includegraphics[scale = 0.5]{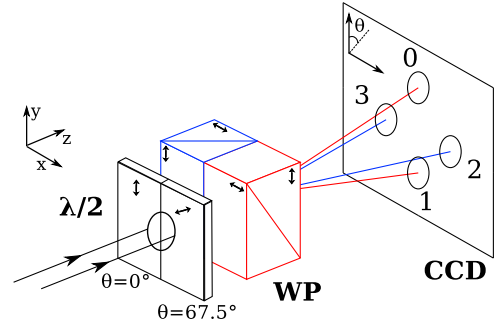}
\caption{Schematic of the half-wave plate ($\rm \lambda /2$) and Wollaston prism (WP) system of RoboPol. The fast optical axes of the elements are shown with bidirectional arrows. The angle of the axis, $\theta$, is denoted below each half-wave plate.}
\label{fig:WP}
\end{figure}

\begin{figure*}
\centering
\includegraphics[scale = 0.6]{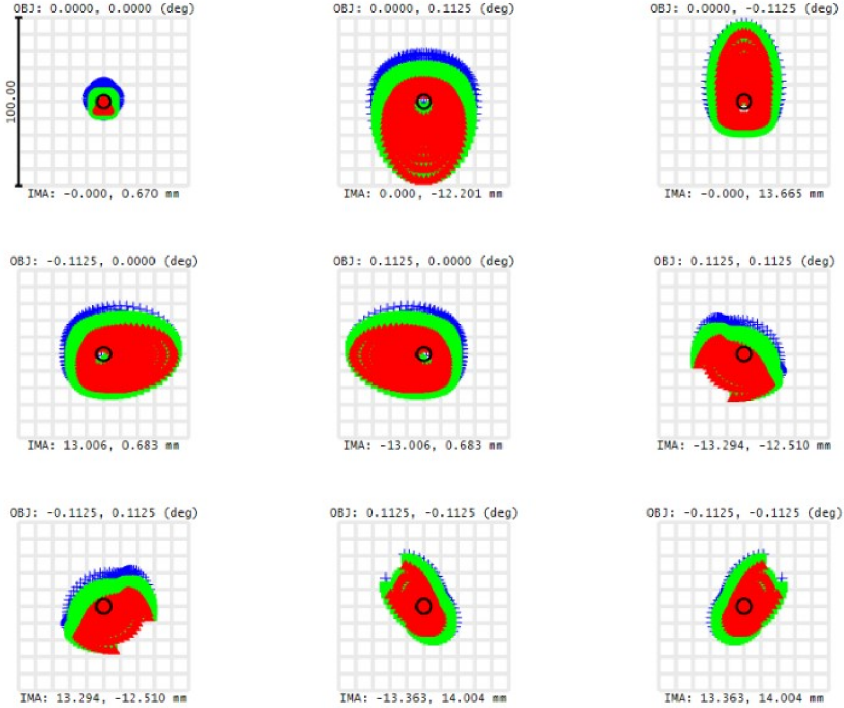}
\caption{Instrument spot diagrams for one of the images formed by one of the RoboPol Wollaston prism $-$ half-wave plate pairs. The image was created using the Zemax software and does not take into account atmospheric seeing. Each panel shows the predicted image for a different position on the FOV. The positions are labeled above each panel. The top left panel shows the image at the center of the FOV, at coordinates (0,0). Other panels show images near the edges of the FOV. Different colors show the image for different input wavelengths: 600 nm, blue, 650 nm, green and 700 nm, red (all within the R band). The scale of the panel (100 $\mu m $, or 2.96\arcsec) is shown next to the top left panel. The CCD pixels have a size of 13.5 $\mu m $ (0.4\arcsec) on the side.}

\label{fig:spot_diagram}
\end{figure*}

\subsection{Optical design} \label{subsec:optical_design}

A side-view of the optical design of RoboPol is shown in Figure \ref{fig:optical_design}. The optical design of the instrument was optimized after including the defocused telescope optics. A combination of lenses collimates the beam from the telescope focal plane and transfers it to the polarization-analyzing system. For simplicity, we only show the rays output from one of the Wollaston prisms. The output beams are imaged on the CCD detector by the camera assembly. The telescope and mount structure place stringent space constraints on the instrument design. As a result, the instrument is very compact, measuring only 421 mm in length.

The polarization optics of the system are shown schematically in Figure \ref{fig:WP}. The telescope beam is shared by a pair of half-wave plates followed by a pair of Wollaston prisms. The angle $\theta$ of the optical axis of the elements is measured clockwise from the y axis as shown on the CCD plane in Figure \ref{fig:WP}. The left half (as shown in Fig. 2) of the beam is transmitted through a half-wave plate with fast axis at $\theta = 0^\circ$. It subsequently passes through a Wollaston prism with axis also at $\theta = 0^\circ$. A light ray transmitted through this pair of elements is split into two rays with orthogonal polarizations: an extraordinary ray (e-ray, with electric field at $\theta = 0^\circ$) and an ordinary ray (o-ray, with electric field at $\theta = 90^\circ$). The o- and e- rays diverge horizontally (along x).  The right half of the beam is transmitted through a half-wave plate with optical axis at $\theta$ = $67.5^\circ$. The subsequent Wollaston prism has its axis at $\theta = 90^\circ$ and splits light rays vertically. Thus, the polarization angle of the beam incident on the Wollaston prism on the right side is effectively rotated by 45$^\circ$ with respect to the prism axis. All elements of the system are fixed and do not move. The specifications of the polarization optics are listed in Table \ref{tab:WPchar}.

Figure \ref{fig:spot_diagram} shows \textit{Zemax}\footnote{www.zemax.com} spot diagrams for modeled point sources placed at different locations within the field of view (different panels). Rays of three different wavelengths (all within the R band) were traced and are shown with different colours. For simplicity, we only show the predictions for one half of the pupil beam passing through a set of half-wave plate and Wollaston prism.

The pattern of divergence of the rays from the origin depends strongly on location in the field of view but only slightly on the wavelength. Any chromaticity induced within the instrument (mainly the WP) is small compared to the size of the typical PSF. The least divergence is seen for a source placed at the center of the field (at (0$^\circ$,0$^\circ$) top left panel), where the maximum divergence is 11 $\mu$m. The highest divergence is 47 $\mu$m, at (0$^\circ$,0.1125$^\circ$). The RMS radial divergence of the rays for a given location on the FOV ranges from 6 $\mu$m to 23.5 $\mu$m.

This RMS radius can be taken as a rough estimate of the PSF size (without the effect of atmospheric seeing). We compare these values to the expected atmospheric seeing. A PSF of 1$\arcsec$ (median seeing at the Skinakas Observatory) is fully sampled with 2.5 pixels (with 0.4\arcsec per pixel on average). This corresponds to a radius of 16.9 $\mu$m, for a pixel size of 13.5 $\mu$m. For the central source, therefore, the PSF is predicted to be seeing-limited. This is not the case for sources placed in the majority of locations on the FOV. In Section \ref{sec:commissioning} we compare these predictions to the actual measured performance of the instrument in terms of encircled energy diagrams.

Differential photometry of the pair of vertical spots gives the relative Stokes parameter $u$, and that of horizontal spots gives $q$\footnote{The normalized Stokes parameters are defined as $q = Q/I$ and $u = U/I$, where $I$ is the total intensity and $Q$ and $U$ are the linear polarization Stokes parameters.}:
\begin{equation}\label{eqn:qu}
q = \frac{N_2-N_3}{N_2+N_3},~~ u = \frac{N_1-N_0}{N_0+N_1},
\end{equation}
where $N_i$ is the photon count of the spot with index $i$ (from 0 to 3) as shown in Figure \ref{fig:WP}\footnote{We note that our equations \ref{eqn:qu} differ from those given in \citet{King2014pipeline} due to a typographical error in the latter.}. The photon counts $N_i$ result after correcting the measured counts for sky background. The uncertainties of the Stokes parameters are given by the following equations \citep[by error propagation, see also][]{Ramaprakash}: 
\begin{equation}\label{eqn:querr}
\sigma_q =  \sqrt{\frac{4(N_{2}^2\sigma_{3}^2+N_{3}^2\sigma_{2}^2)}{(N_{3}+N_{2})^4}},~~ \sigma_u = \sqrt{\frac{4(N_{0}^2\sigma_{1}^2+N_{1}^2\sigma_{0}^2)}{(N_{0}+N_{1})^4}},
\end{equation}
while the uncertainties of the intensities, $\sigma_i$, are calculated according to \cite{Laher2012}:
\begin{align}
\sigma_i = & \sqrt{N_i + \sigma^2_{\rm sky}A_{\rm phot}+\frac{\sigma^2_{\rm sky}A_{\rm
phot}^2}{A_{\rm sky}}},
\end{align}
where $\sigma^{2}_{\rm sky} = n_{\rm sky}$ is the average sky intensity
(background) in a single pixel, $A_{\rm phot}$ is the area (in pixels) of the photometry aperture, and $A_{\rm sky}$ is the area of the annulus used for background estimation. The first two terms account for the photon counting statistics of the source and sky, and the third describes the background estimation uncertainty.

\begin{figure}
\centering
\includegraphics[scale = 0.5]{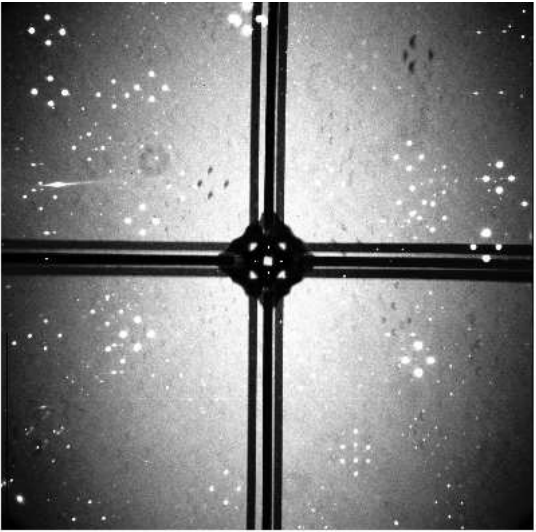}
\caption{An exposure taken with RoboPol. The colors have been stretched to highlight features in the FOV. Background light in the central region is partially blocked using a focal-plane mask. The dark vertical and horizontal lines are the shadows cast by the mask supports. There is significant vignetting throughout the FOV, resulting in large-scale variations of the sky background intensity. Small-scale artifacts are also present and are due to the presence of dust specs at different locations within the instrument.}
\label{fig:rbpl_img}
\end{figure}

The design of RoboPol\footnote{The authors are happy to share more design details through private communication.} differs from that of most four-channel polarimeters. It uses a half-wave plate to rotate one half of the incoming light instead of a modified Wollaston prism such as that proposed by \citet{Geyer1996}. In this respect it is similar to the design described in \citet{fujita2009}. However, \citet{fujita2009} used a beam splitter to divert half of the beam to one of the prisms, and channeled that part of the beam on the detector by use of a folding mirror. RoboPol's design, in contrast,  avoids the large instrumental polarization induced by a folding mirror.

Another difference with other four-channel designs is that the Wollaston prisms do not contain wedges such as those proposed by \citet{oliva1997} and implemented in many instruments \citep{pernechele2003,kawabata2008,afanasiev2012,covino2014,helhel2015,devogele2017}. As a result, there is significant vignetting throughout RoboPol's FOV (Figure \ref{fig:rbpl_img}). Additionally, there are well-described geometric distortions that cause the 4-spot pattern to vary as a function of position. These effects are taken into account in the modeling of the instrumental response described in section \ref{sec:commissioning}.

\begin{table}
    \centering          
    \caption{Wollaston prism and half-wave plate characteristics.}             
    \label{table:WP}      
    \begin{tabular}{c  l }   
      \hline
      Wollaston Prisms&\\
      \hline                     
       Material & Quartz \\ 
       Size 	& 25$\times$25 mm	\\	
       Clear Aperture&  22.5 mm minimum\\
       Divergence& 60$\arcmin$ in the visible\\
       Wavelength Range& 400-900 nm \\
       Extinction Ratio& < 10$^{-4}$ \\
       Wavefront Distortion& $<\lambda$ at 633nm\\
       AR coating& $R$< 0.7\% over 500-900 nm\\& on both surfaces\\
       Manufacturer & Karl Lambrecht corp.\\
       \hline   
       Half-wave plates  & \\
       \hline
       Material & MgF2 and Quartz crystal\\ &(cemented)\\
       Retardation& $\lambda/2 \pm 5\%$ over 400-900 nm \\
       Beam Deviation& $< 1\arcmin$\\
       Wavefront Distortion& $< \lambda$ at 633nm\\
       Accuracy of axis orientation& $\pm 30\arcmin$\\
       Size & 25$\times$25 mm\\
       AR coating & $R < 0.5 \%$ over 500-900 nm \\& on both surfaces\\
       Manufacturer &  Karl Lambrecht corp.\\
       \hline
    \end{tabular} 
    \label{tab:WPchar}
  \end{table} 
  
\subsection{Mask}

\begin{figure*}
\centering
\includegraphics[scale = 1]{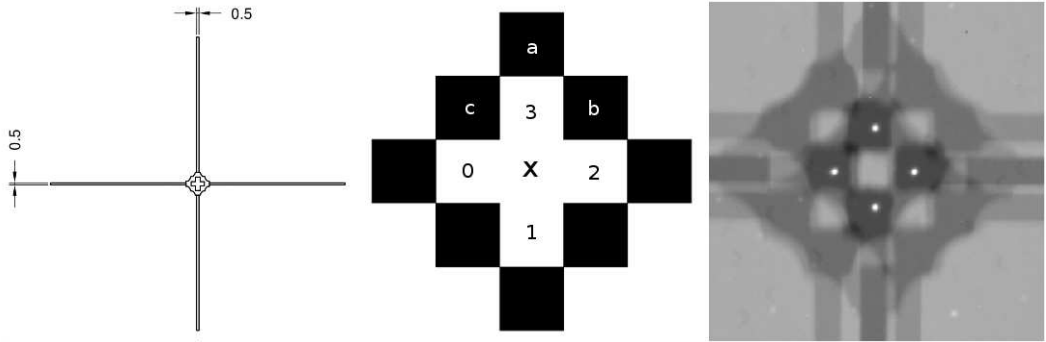}
\caption{The RoboPol mask. Left: Design of mask and its supports, with the physical dimensions  marked (in mm). Middle: Schematic of mask shape. White indicates open space while dark indicates the presence of light-blocking material. Right: Zoom-in of a typical RoboPol science image showing a point source placed in the darkest shadows of the mask in the center of the field of view. The linear shadows extending towards all edges of the image are cast by the supports of the mask. }
\label{fig:mask}
\end{figure*}

The four-channel design of RoboPol has two disadvantageous consequences due to the fact that any point on the CCD receives photons from four different regions of the sky:
(a) the photon background (noise) is increased compared to (e.g.) dual beam polarimetry and (b) neighbouring sources on the sky can overlap as projected on the CCD. To address these issues, it is customary in four-channel polarimetry to place a mask that blocks light from nearby regions of the observed target field, reducing the sky background \citep[e.g.][]{kawabata2008}. We followed this approach in order to increase the measurement accuracy, but only for the central target. 

Figure \ref{fig:mask} (left) shows the mask design and its supporting legs. The middle panel shows a schematic of the mask (center of left panel). Open (unblocked) regions are colored white, while the areas that are blocked by the mask are colored black. The light from the central square (marked 'x') is projected on the $\sim 21\arcsec$-wide squares marked 0-3. Square 3 receives light from the central square ('x') but not from squares a, b, and c. It receives one quarter of the light reaching the central square. The remaining three quarters of the light from the central square are divided among regions 1, 2, and 0. As a consequence, the background for the central source is reduced by a factor of $\sim$four compared to the majority of the field of view. The right panel of Fig. \ref{fig:mask} shows the shadows cast by the mask during a science exposure with RoboPol.

The support structure of the mask can be seen extending in four directions in Figures \ref{fig:rbpl_img} and \ref{fig:mask}. Along with the mask, these supports cast shadows, reducing the available for polarimetry area to $\sim$85\% of the total ($13.6\arcmin \times 13.6\arcmin$) field of view. The mask appears unfocused as it is not located exactly at the focal plane. This was necessitated due to the need to contain the instrument within the space available.

The design of the mask has to take care of two important considerations. First, reflection from the mask surface can impart erroneous polarization in the light from the astronomical source. Therefore, the mask surface was coated with plastic material of 10 to 15 $\mu$m thickness on each surface to become non-reflective. Second, the size of the mask was selected to simultaneously (a) minimize photon contamination from the neighborhood of the central target and (b) allow for enough area for background estimation around the target. 

The inner portion of the mask (depicted in Fig. \ref{fig:mask}) measures 5.2 mm $\times$ 5.2 mm $\times$ 1 mm, and the full size including the supporting legs is 66.2 mm $\times$ 66.2 mm $\times$ 2.5 mm. 

\subsection{Control System and Data reduction Pipeline} \label{subsec:control}

One of the design goals of the RoboPol instrument was to
operate with high observation efficiency, which was achieved
by fully automating the observing process. The RoboPol control system runs the telescope on robotic mode when the
instrument is online, but allows the telescope to be run manually when the instrument is not in use. The control system features automated target acquisition, telescope focusing, dynamic exposure time calculation, and target of opportunity observations (including an alert system for GRBs).

The large amount of data generated by the survey is handled by an automated data reduction pipeline, specifically developed for this purpose. The program performs aperture photometry of all point sources in the field of view, calculates their normalized Stokes parameters $q$ and $u$ and also provides relative photometry. A detailed description of the RoboPol control system and pipeline is given in \citet{King2014pipeline}. Upgrades made to the pipeline, mainly with regards to analysis in the wider FOV, are described in \citet{Panopoulou2015}.

By default, the data are reduced and corrected for the instrumental polarization according to the instrument model described in \citet{King2014pipeline} (see also section \ref{sec:commissioning}). The parameters of the model are fit anew each observing season, as the instrument is removed from the GAM at the end of the observatory seasonal operations (typically November) and re-installed at the start of the following season (typically March). A separate model is constructed for each filter.

\begin{table}
    \centering          
    \caption{Literature polarization of standard stars used for instrument
	     calibration.}             
    \label{table:standards_table}      
    \begin{tabular}{c c c c l l l }   
      \hline\hline                     
       Name& $p$(\%) & $\chi(\circ)$ & Band & Ref \\ 
       \hline  
       BD +32 3739 	& 0.025$\pm$0.017	& 35.79$^{\circ}$		& V &	1	\\
       G191B2B	& 0.061 $\pm$ 0.038	& 147.65$^{\circ}$      & V & 	1\\
       HD 212311 	& 0.034 $\pm$ 0.021 	& 50.99$^{\circ}$   	&  V & 1\\
       HD 14069	& 0.022 $\pm$ 0.019  	& 156.57$^{\circ}$   		&  V & 1\\
       BD +59 389 	& 6.430$\pm$0.022	& 98.14$^{\circ}$ $\pm$ 0.10$^{\circ}$& R &	1	\\ 
       BD +33 2642   & 0.20$\pm$0.15  & 78$^{\circ}\pm$ 20$^{\circ}$  &R & 2\\        
       WD2149+021   & 0.05  & -125$^{\circ}$  &R & 3\\ 
       HD 154892 &$0.05\pm0.03$& -- &B&4\\
       BD +40 2704& $0.07\pm0.02$& 57$^{\circ}\pm 9^{\circ}$& ?& 5\\
       \hline          
    \end{tabular} 
    (1)\citet{schmidt}; (2) \citet{Skalidis2018}; (3) \citet{cikota2017}; (4) \citet{Turnshek}; (5) \citet{Berdyugin2002}
    \label{tab:standards}
  \end{table} 
  
\section{Commissioning} \label{sec:commissioning}

\begin{figure}
\centering
\includegraphics[scale =1]{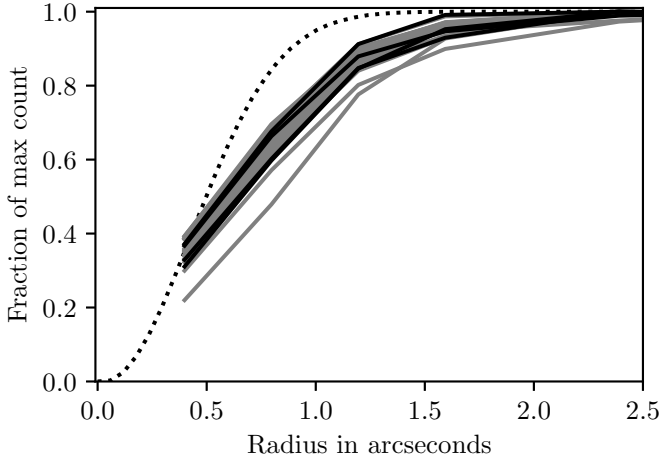}
\caption{Fraction of total energy that is enclosed within an aperture of given radius (growth curve). The dotted line is the predicted curve from the Zemax software for a source at the center of the field (there is no significant difference between predictions for different spots of the same source). Solid gray lines show the observed growth curves for 6 sources located throughout the FOV (a separate curve is shown for each of the four spots), while the solid black lines are for a source in the center of the field (within the mask).}
\label{fig:growth}
\end{figure}

The RoboPol instrument was designed, assembled and tested at IUCAA before shipping to the Skinakas observatory for commissioning in May 2013. The mechanical and optical elements of the design were found to be in excellent working order. Slight modifications were made to the telescope weight distribution. Well-known polarization standard stars were observed a multitude of times to validate the performance of both the instrument and the data reduction pipeline.

During commissioning, a model of the instrumental response throughout the FOV was developed, as explained in \citet{King2014pipeline}. Unpolarized standard stars were used to raster map the FOV. These observations were then used to develop an instrument model; i.e. a set of functions that describe the variation of (a) the spot pattern, (b) the total intensity and (c) the instrumental Stokes parameters across the FOV. The residuals resulting from subtraction of the model from the data are uniform across the field \citep[see Fig. \ref{fig:Rmodel} and also][]{King2014pipeline}, testifying the effectiveness of the model to remove systematic large-scale patterns in the aforementioned parameters. The model is agnostic as to what optical effects it corrects for (e.g. vignetting, half-wave plate non-uniformity). All large-scale systematic effects expected to affect polarization measurement to the required accuracy are modeled out with this approach.

The on-sky characteristics of the PSF were compared to those predicted during the design phase. This comparison is made in terms of the fraction of total energy of a source that is enclosed within a circular aperture of given radius (curve of growth). We show a comparison between the curve of growth predicted from the Zemax model and those resulting from observations in Fig. \ref{fig:growth}. In order to minimize the effect of varying atmospheric seeing on this comparison, we have selected 6 sources observed during a night with median seeing 1\arcsec. The exposure time was 20 seconds and the R band magnitudes of the sources were in the range 12$-$16 mag. We show a separate curve of growth for each of the four spots of a source. We find significant differences between the observed and predicted curves. First, the observed curves of growth reach 90\% of the total light at a radius of 1.2\arcsec$-$1.6\arcsec, which is 1.3$-$1.8 times larger than the radius expected from the Zemax model (0.9\arcsec). Second, the observed curves exhibit differences between the vertical and horizontal spots, something that is not found by the model. In particular, the two extreme curves that rise less steeply in Fig. \ref{fig:growth} correspond to the horizontal spots of a source.

Such differences from the Zemax prediction are to be expected, as the prediction does not take into account a multitude of factors which are present in realistic situations (e.g. optical element misalignment, telescope tracking jitter, imperfect seeing conditions). In practice, we take care to perform photometry within apertures that enclose the majority of the energy of a source \citep[e.g.][]{Panopoulou2015}, and demonstrate that this results in excellent instrument performance (see Section \ref{sec:performance}).

An addition made to the mechanical system was the introduction of a pump used to channel dry air onto the glass protecting the CCD to prevent water vapor condensation on its surface during nights with high humidity. Finally, a removable plastic cover was placed on the window of the instrument, during the time that the telescope was in the stow position, to prevent dust from settling on the surface of the first lens.

\begin{figure}
\centering
\includegraphics[scale = 0.6]{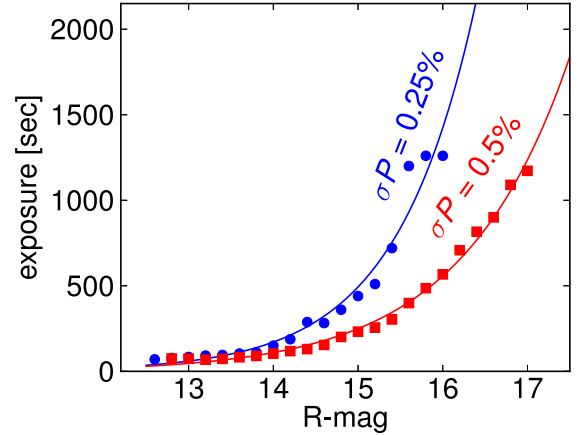}
\caption{Required exposure time as a function of R-band magnitude in order to reach a $\sigma_p$ of 0.25\% (blue) and 0.5\% (red). Squares and circles represent median values for multiple measurements obtained at different elevations, atmospheric conditions and Moon phase. Curves show the exponential function fit to the data: $a\,e^{bx}+c$, with ($a$, $b$, $c$) = (3.2$\times10^{-5}$, 1.09, 39) and (0.0018, 0.79, -6) for a target $\sigma_p$ of 0.25\% and 0.5\%, respectively.}
\label{fig:exptime}
\end{figure}

\begin{figure*}
\centering
\includegraphics{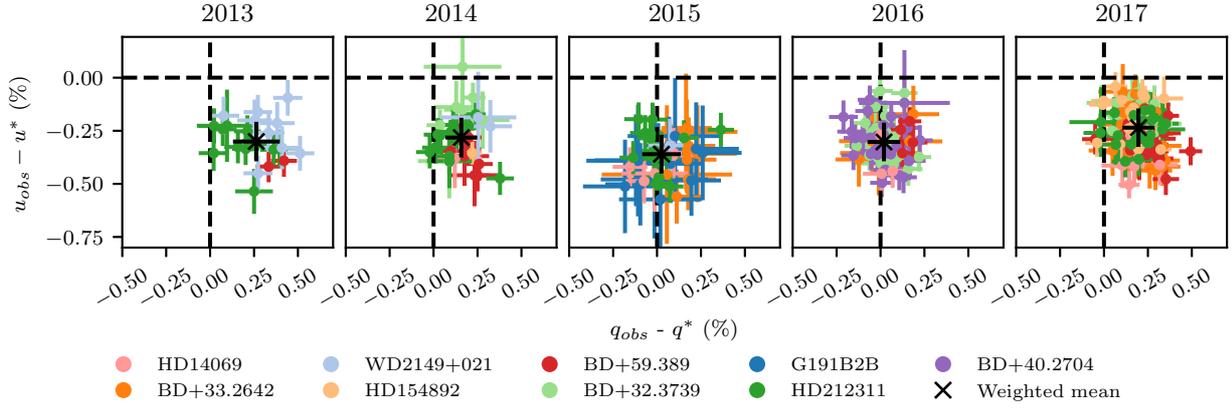}
\caption{Standard star residuals from literature values (shown in Table \ref{tab:standards}) for all years of RoboPol operation. Each point is a single observation, processed without using the instrument model. A different color is used for each star. Observations were conducted in the mask and in the R-band. Error bars shown are only statistical. The black cross marks the weighted mean of all measurements. Dashed lines are for visualization of (0,0).}
\label{fig:allqu}
\end{figure*}

\begin{figure*}
\centering
\includegraphics{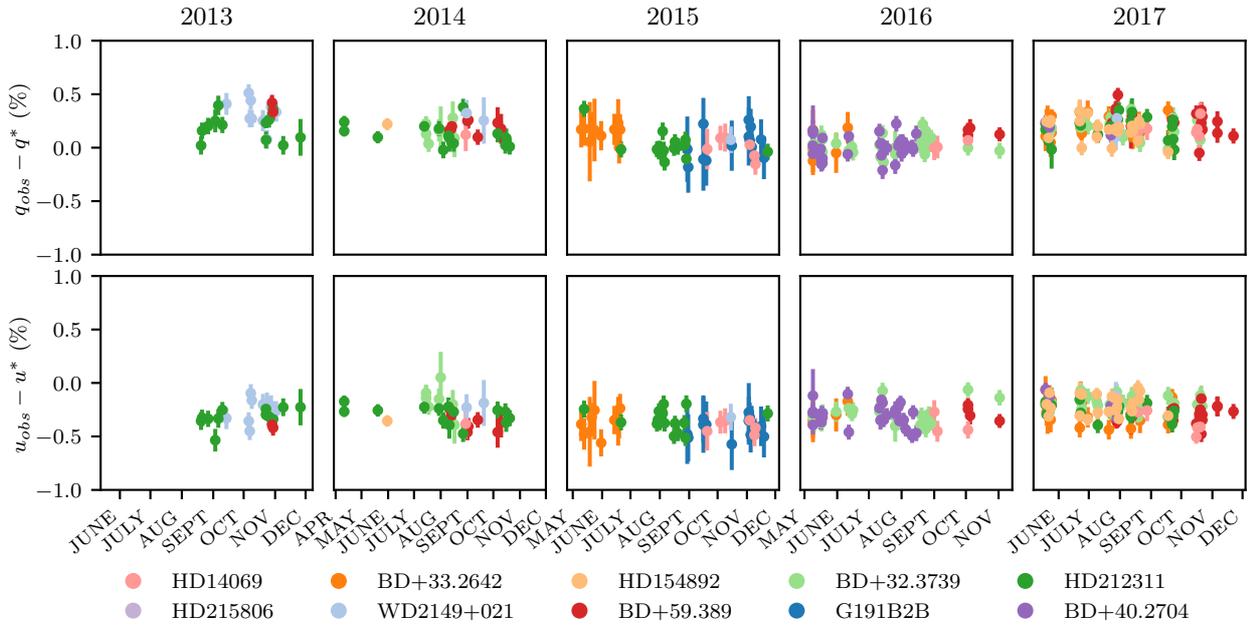}
\caption{The same measurements of Figure \ref{fig:allqu} but shown as a function of observing date for all years of RoboPol operation.}
\label{fig:qujd}
\end{figure*}

\section{Performance} \label{sec:performance}

In the five years of RoboPol operations, the efficiency and accuracy of the instrument have surpassed design specifications. In terms of time efficiency, the combination of the instrument design (no moving parts) with the automated observing strategy have resulted in an average of 15 targets observed per night during the 3-year monitoring programme.

\begin{figure}
\centering
\includegraphics[scale = 1]{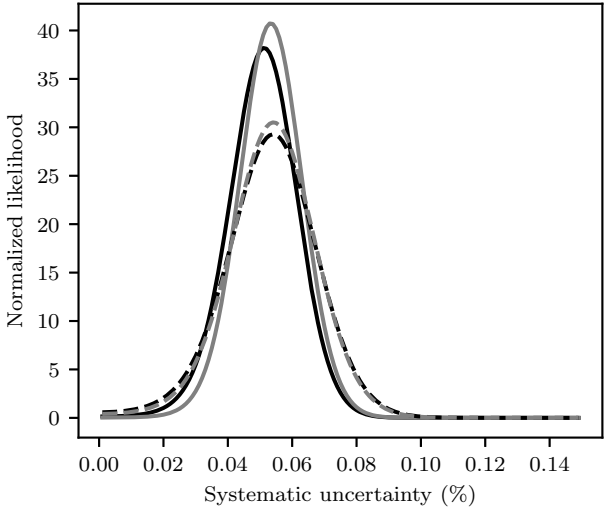}
\caption{The normalized likelihoods for the systematic uncertainty of $q_{\rm inst}$ (black) and $u_{\rm inst}$ (gray) in the mask, calculated using measurements of standard stars (Fig. \ref{fig:allqu}) from 2016 (dashed) and 2017 (solid). The maximum-likelihood systematic uncertainty is $\sigma_{\rm inst} = 0.051-0.054$ \%.}
\label{fig:s0}
\end{figure}

We characterized the performance of the instrument in terms of exposure time using observing data of different sources placed within the mask throughout all observing seasons. The data used cover different elevations, atmospheric conditions and Moon phases. In Fig. \ref{fig:exptime} we present the required exposure time in order to reach 0.25\% and 0.5\% (statistical) accuracy in polarization for sources of different magnitude.

\subsection{R-band performance within the mask}
\subsubsection{Instrumental polarization}
During the five years of operation, standard calibrator stars were observed each night along with the science observations. These 
measurements allow us to evaluate the instrumental polarization and its uncertainty (systematic uncertainty) as a function of time. For our initial analysis we do not make use of the instrument model presented in Section \ref{sec:commissioning}. The literature values of standard stars used for calibration are shown in Table \ref{table:standards_table}.

The instrument causes the observed Stokes parameters of standard stars, $q_{\rm obs}$ and $u_{\rm obs}$, to be offset from their literature values, $q^*, u^*$, in a systematic way. In other words, the literature-corrected measurements $\bar{q} = q_{\rm obs} - q^*$ and 
$\bar{u} = u_{\rm obs} - u^*$ are found to be offset from (0,0) on the $\bar{q}-\bar{u}$ plane. Measurements of standard stars are shown on the $\bar{q}-\bar{u}$ plane in Figure \ref{fig:allqu}, grouped by year of observation. Each point is a single observation, with errors that are purely statistical (from photon noise). All observations have been processed by the RoboPol pipeline, \textsl{without} making use of the instrument model. This allows us to determine the systematic uncertainty due to the instrument alone, avoiding possible unknown errors due to modeling. The pipeline version used employs optimized aperture photometry as described in \citet{Panopoulou2015} with slight modifications presented in \citet{Skalidis2018}.

For all years, the instrument introduces a fractional linear polarization at the level of $p_{\rm inst} = 0.3-0.4\%$. This is found using the weighted mean of all $\bar{q}$ and all $\bar{u}$ measurements (black crosses in Figure \ref{fig:allqu}). The level of $p_{\rm inst}$ varies by less than $0.1\%$ throughout five years of observing, during which there have been multiple removals and replacements of the instrument on the telescope.

The scatter of $\bar{q}$ and $\bar{u}$ measurements contains information on the level of random variation of the instrumental polarization. Normally this scatter is also influenced by a number of other factors: observational uncertainties in the measurements of the standard stars, atmospheric variations throughout the observing period, errors in the literature values of standards, and possible intrinsic variability of standards. Seeing has been found to significantly affect measurements of standard stars with a nearby source (within a few arcseconds). One example is Hiltner 960. This is because the second source is (partially or fully) blended with the calibrator star during nights with bad seeing. \citet{slowikowska2016} find a similar problem with a source that lies 16\arcsec \, away from BD+59 389 using the RINGO3 polarimeter. We have found this to not be the case for our measurements, as the sources are well-separated for all the nights observed. Note that even though commonly used standard stars are assumed to be stable, we have found this not to be the case for a subset of regularly monitored stars. The wealth of data collected on standard stars throughout five years of operation, along with the aforementioned stability of the instrument, has allowed us to identify a subset of standard stars that are variable. We have also found a number of stars that appear offset from the rest of the calibrators in the $\bar{q}-\bar{u}$ plane (and therefore have erroneous literature values). These stars are not included in this analysis, and we will dedicate a separate publication to present and discuss them (Blinov et al., in prep).

In this analysis we only make use of standards that we trust have no or undetectable (below $0.1\%$) intrinsic variability and have accurate literature values (these are the stars used for Fig. \ref{fig:allqu} as well). We investigate whether there are detectable variations of the polarization of standards within a season in Figure \ref{fig:qujd}, which shows the same observations as in Figure \ref{fig:allqu} as a function of time. The lack of points in 2013 is due to the fact that very few high-quality standards were observed initially. As time progressed, we refined our set of observed standards. We also increased exposure times (resulting in reduced errors for the 2016, 2017 set) and frequency of observations within a given night.

In order to de-couple these effects from the instrumental polarization variability, we adopt a novel approach. First, we make use only of the aforementioned well-behaved standards. We then make the assumption that the instrumental variability in $q$ and in $u$ follows Gaussian distributions and we treat $q$ and $u$ independently. The errors on the measured $\bar{q}$, $\bar{u}$ are Gaussian (photon signal-to-noise ratios are of order $10^3$). In this case, the likelihood function for the instrument variability can be found analytically, and is given by equation A5 in \citet{2007ApJ...666..128V}. 

We calculate the normalized likelihoods for the systematic uncertainty of the instrumentally-induced normalized Stokes parameters $q_{\rm inst}$ and $u_{\rm inst}$ separately, using measurements obtained in a single observing year with RoboPol (Figure \ref{fig:s0}). The maximum likelihood scatter of the instrument polarization is found to be $\sigma_{\rm inst} = 0.051 - 0.054\%$ for both $q$ and $u$. Values within this range are found when using the observations of 2016 (74 measurements) and of 2017 (130 measurements). However, for smaller numbers of observations the error in the maximum likelihood estimate will be larger. This is the case for the years 2013, 2014 and 2015, where only 23, 33 and 47 measurements can be used for the determination of the instrumental polarization. In these initial seasons, many standards were observed that were in fact variable. Some of these stars were used for the initial determination of the instrument performance \citep{King2014pipeline}.

It may also be argued that part of the systematic variability measured could arise from the fact that we are using observations throughout an entire season (May-November) for its determination. This is not the case, however, as we do not find a significant shift in the $\bar{q}$ or $\bar{u}$ measurements at different dates within a season, as evidenced by inspection of Fig. \ref{fig:qujd}.

We now explore how the introduction of the instrument model may change the above conclusions. To this end, we processed the 2017 set of standards using the model built in the same year (using the star HD 212311). As expected, we find the residuals to be reduced: the weighted mean $\bar{q}$ is -0.07\% and that of $\bar{u}$ is 0.017\% (compared to 0.18 and -0.24 \% without the model correction). We find that the maximum likelihood scatter in the instrument polarization in 2017, after applying the model is $\sigma_{\rm inst,q}$ = 0.054\%, $\sigma_{\rm inst,u}$ = 0.052\%, consistent with that found without using the model correction (0.051\% and 0.053\%, respectively). 

\subsubsection{Instrumental rotation}
Aside from the offset on the $\bar{q}-\bar{u}$ plane, another instrumental effect is the a rotation of the instrument frame compared to the celestial frame. We can measure this rotation by using polarized standard stars, which have known and well-measured polarization angles ($\chi^*$). In order to measure the rotation, we first correct each measurement of a polarized standard for the instrumental zero-point offset found previously. In practice, we subtract $q_{\rm inst}$ and $u_{\rm inst}$ from each measurement of the polarized standard and propagate the errors. Then, we find the (corrected for zero-point-offset) polarization angle $\chi_{\rm obs,c}$ and subtract from it the literature value $\chi^*$. Any deviation from 0 points to an instrument frame rotation (compared to the sky). 

\begin{figure}
\centering
\includegraphics[scale = 1]{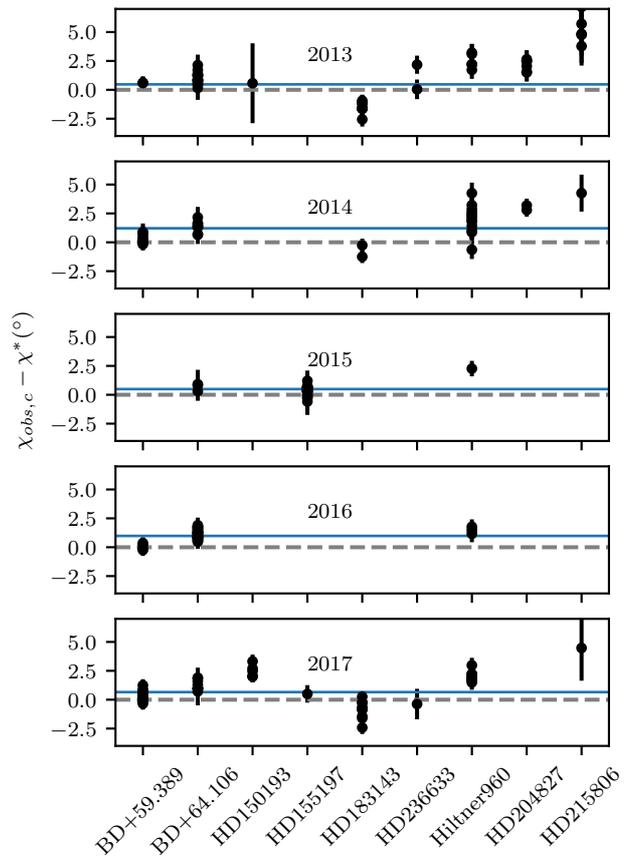}
\caption{Rotation of the instrument reference frame compared to the celestial frame measured with polarized standard stars.  All years from 2013 (top) to 2017 (bottom). The dashed line marks a rotation of 0$^\circ$. The mean rotation for each year is marked by the solid blue line.}
\label{fig:rotang}
\end{figure}

These differences are shown in Figure \ref{fig:rotang}, for all years of RoboPol operation\footnote{Literature values for the polarized standard stars in Fig. \ref{fig:rotang} were taken from: \citet{schmidt} for BD+59.389, BD+64.104, HD 155197, HD 236633, Hiltner 960, from \citet{whittet1992} for HD 150193, HD 215806, and from \citet{hsu1982} for HD 183143, HD 204827}. There is considerable scatter not only between observations of different stars but also between the measurements of an individual star compared to the errors. This is most likely a result of the variable nature of the majority of these standards (HD183143, HD204827, \citet{bastien1988}, HD155197, HD236633, Hiltner960, \citet{schmidt}, HD150193, \citet{Hubrig2011}). Their variability excludes them from being used to measure the zero-point-offset. However, because they are highly polarized, they can still serve for estimating the instrumental polarization angle rotation. Another source that could be contributing to the observed scatter is the variation of the instrument coordinate system that results from removal and repositioning of the instrument on the telescope (which happens up to 2-3 times throughout an observing season). With the existing dataset, we cannot distinguish between these two sources of uncertainty. However, even with the uncertainties introduced, we find the weighted mean instrumental polarization angle rotation (solid blue line) to be 0.5$^\circ< \chi_{\rm inst}<1.2^\circ$ for all years. 

For the year 2013, the instrument rotation found here is smaller than that found in \citet{King2014pipeline} ($2.31^\circ \, \pm \,0.34^\circ$). This difference arises mainly from the fact that we do not make use of the standard star VI Cyg \#12 \citep{schmidt} in our analysis (which shows signs of strong variability in our data), and include the star HD 183143. If we use the same sample as \citet{King2014pipeline}, we find values consistent within 1$\sigma$. This underscores the necessity of establishing a large set of reliable polarized standards, to allow a precise determination of the instrument reference frame rotation.

\subsection{R-band performance throughout the FOV}
\begin{figure}
\centering
\includegraphics[scale = 1]{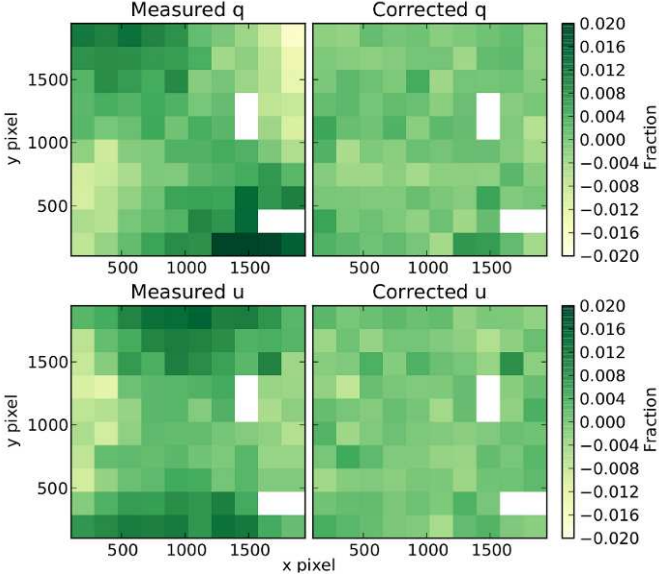}
\caption{Comparison between $q$ and $u$ measurements of an unpolarized standard star positioned throughout the field: left, without model correction and right, with model correction. White squares are due to the absence of measurements in those positions.}
\label{fig:Rmodel}
\end{figure}

As discussed in section \ref{sec:instrument} the instrumental response is a function of target position on the FOV. To characterize this response for the entire FOV, we make use of the instrument model. As demonstrated in \citet{King2014pipeline}, the instrument model is capable of removing the large-scale patterns seen in the instrumental $q$ and $u$. Figure \ref{fig:Rmodel} shows R-band $q$ and $u$ measurements of the unpolarized standard star HD 212311, which was used to create the model of 2017. The instrumental polarization prior to model correction is shown in the left panels (top for $q$, bottom for $u$). The residuals after model correction (shown in the right panels) are spatially uniform and lie below a level of $0.3\%$. The systematic uncertainty outside the mask remains at these levels for models taken in different years of RoboPol operation \citep[compare with][]{King2014pipeline,Panopoulou2015}.

Apart from the large-scale spatial variations in the instrument response, there exist small-scale features that affect the instrumental polarization locally. These small-scale features are due to the presence of dust particles that lie within the instrument. The particles cast shadows in different positions in the FOV, as can be seen in Figure \ref{fig:rbpl_img}. Due to the design of RoboPol, these features cannot be simply corrected for by flat-fielding. Our approach is to detect these features in flat-field images (taken with the telescope pointed at the sky during twilight), and then discard any targets that happened to be observed on the position of a dust spec. This procedure is explained in detail by \citet{Panopoulou2015}. 

\subsection{Instrument characterization in B,V and I bands}

\begin{figure}
\centering
\includegraphics[scale = 0.9]{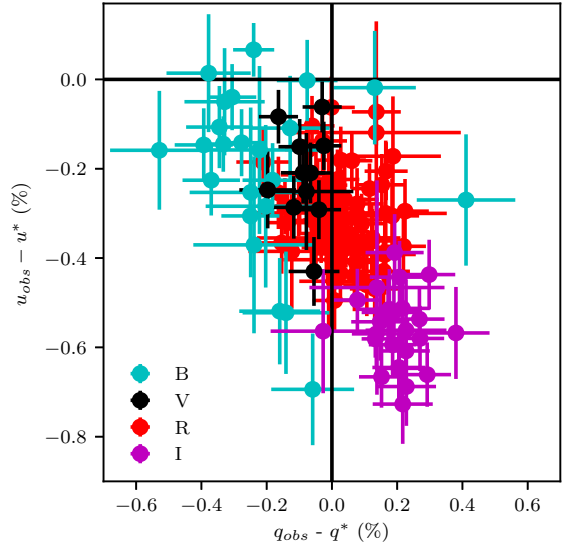}
\caption{Comparison between $q$ and $u$ residuals of standard stars in the mask (from their literature values) observed during 2016 in the B, V, R and I bands. The measurements have not been corrected for instrumental polarization (using the model).}
\label{fig:bvriqu}
\end{figure}

The majority of observations are performed in the R-band. However, some of the other RoboPol projects require multi-wavelength measurements. Due to the relative scarcity of standard star observations compared to the R-band, we cannot perform a similarly rigorous characterization of the instrument in the B, V and I bands. For this reason, we consider data from the year 2016 within which the most observations of standard stars in these bands were taken. Figure \ref{fig:bvriqu} shows $q$ and $u$ measurements of standards observed in the B, V, R and I bands in the mask. No model correction has been applied to the data. We find that the mean instrumental polarization in the mask varies within 0.6\% between bands. The weighted mean $p_{\rm inst}$ in the different bands are: 0.29 $\pm$ 0.16 \% (B), 0.21 $\pm$ 0.099 \% (V), 0.30 $\pm$ 0.091 \% (R), 0.6 $\pm$ 0.077 \%(I), where the quoted uncertainty was calculated from the standard deviation of $\bar{q}$, $\bar{u}$ measurements.

\begin{figure}
\centering
\includegraphics[scale = 1]{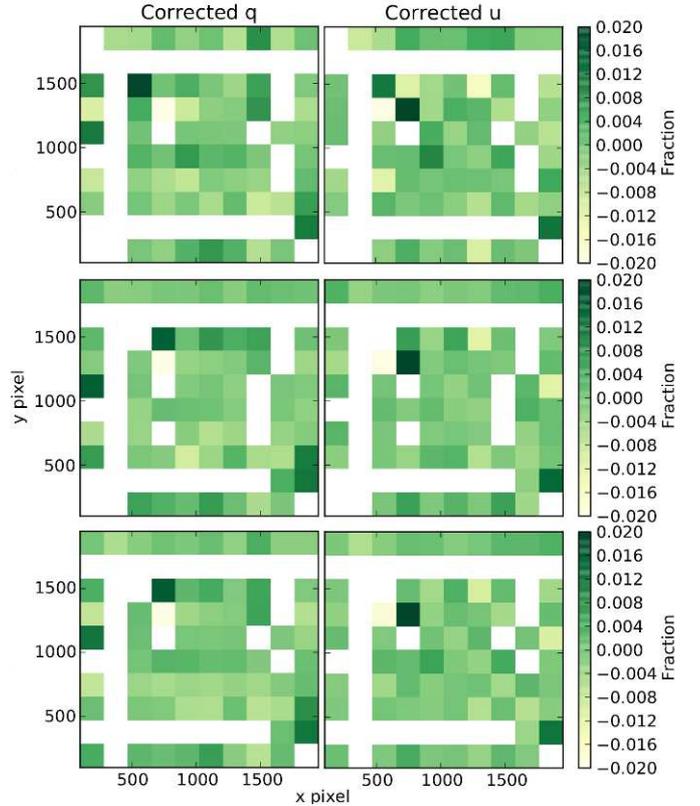}
\caption{Residuals instrumental $q$ (left) and $u$ (right) after model correction throughout the field of view. From top to bottom: B, V, and I band.}
\label{fig:BVImodel}
\end{figure}

For each band, a separate model is created in each observing year. We explore the effectiveness of the model in Figure \ref{fig:BVImodel}. The panels (B, V, I from top to bottom) show the residuals after model correction for 2017. The residuals are spatially uniform, as is the case for the R-band model. In all three bands the rms residuals are at the same level as in the R-band. There are a number of positions that are empty, as a result of poor sampling of the FOV during production of the raster map. This would only affect the analysis of targets outside the mask (falling in the regions with gaps). However, all multi-band observations have placed targets in the mask, where the instrumental polarization is best understood.  
The presented models have been adequate for the purpose of removing of instrumental polarization in the mask. 

Several artificial effects appear in the B and I bands, which render the task of controlling systematics outside the mask more difficult than in the R band.

\begin{figure}
\centering
\includegraphics[scale = 1]{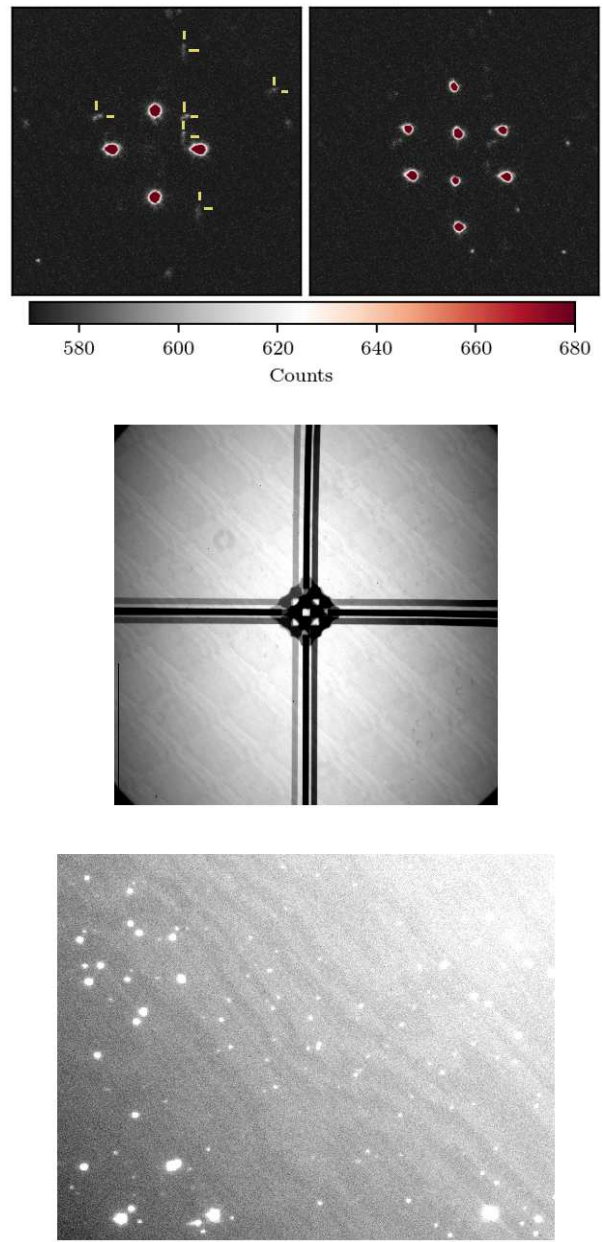}
\caption{Artifacts in the B and I bands. Top: B band ghosts around bright stars. Left: a source in the mask. Each ghost is marked with a vertical and horizontal yellow line. Right: a pair of sources in the field. Middle: Sky flat field taken in the B band. The colors have been stretched to highlight the periodic patterns seen throughout the FOV. Bottom: Zoom in to a region of the FOV where fringes are visible in the I band (exposure time was 2 min).}
\label{fig:defects}
\end{figure}

One artifact that occurs in the B band is the appearance of ghost images near relatively bright stars. An example of such images is shown in Figure \ref{fig:defects} (top) for a source placed in the mask (left) and for a pair of sources in the field (right). The exposure time was 5 seconds and the central source has an R magnitude of $\sim 9$. In the case of the source in the mask, the background sky is too faint for the shadow of the mask to be clearly visible (as a result of the short exposure time). The pattern of the ghosts is similar in both cases, but the ghost images are brighter for the source in the mask (these brighter ghosts are marked with yellow lines in the Figure). This artifact appears to be the result of increased reflectivity of the anti-reflection coating on the CCD window.

A second effect in the B band is the appearance of a periodic striped pattern that runs diagonally throughout the FOV. This is best seen in the flat-field image of Figure \ref{fig:defects} (middle). The intensity variations caused by this pattern are of order 1\%. This effect is also seen in images taken with the same camera, but without RoboPol, and hence is not related to the instrument.

A final artifact is seen in the I band, where a pattern of fringes appears in the background far from the center of the field. These I-band fringes are quite typical of thinned back illuminated CCDs \citep[e.g.][]{2012PASP..124..263H}. They are caused by thin film interference effects for light of longer wavelengths, between the various CCD layers that result in quantum efficiency variations in CCD pixels. An example sub-field of the FOV that exhibits fringing is shown in Fig. \ref{fig:defects} (bottom). The reduction in brightness within the fringes is 1-2\%.

\section{Summary} \label{sec:conclusion}

We have presented the RoboPol four-channel imaging polarimeter, developed for use at the Skinakas observatory 1.3 m telescope in Crete, Greece. It has been operating since it was commissioned in 2013. The main task of RoboPol has been to monitor the linear polarization of a large sample of blazars in the R-band from 2013 to 2015, as part of the RoboPol programme. RoboPol has delivered science for a number of other projects, including Be/X-ray binary and interstellar medium studies.

The design of RoboPol makes use of two Wollaston prisms and two non-rotating half-wave plates to produce simultaneous measurements of the Stokes $q$ and $u$ parameters. 

The R-band performance of RoboPol is stable throughout an observing season and varies very little during five years of regular operation. The scatter in the offset (or instrumental polarization) for R-band measurements in the central mask region is below 0.1\% (0.05\% maximum likelihood value) in fractional linear polarization. Measurements can be performed for point sources throughout the 13.6$\arcmin \times 13.6 \arcmin$ FOV in the R-band where systematic offsets are controlled at the level of 0.3\%.

\section*{Acknowledgments}
We thank Anna Steiakaki for her invaluable contribution and technical support; John Kypriotakis for helping with edits in the paper; and the anonymous reviewer for their suggestions.

The RoboPol project is a collaboration between Caltech in the USA,
MPIfR in Germany, Toru\'{n} Centre for Astronomy in Poland, the University of Crete/FORTH in Greece, and IUCAA in India.
The U. of Crete group acknowledges support by the ``RoboPol'' project, which is implemented under
the ``Aristeia'' Action of the  ``Operational Programme Education and Lifelong Learning'' and is
co-funded by the European Social Fund (ESF) and Greek National Resources, and by the European
Comission Seventh Framework Programme (FP7) through grants PCIG10-GA-2011-304001 ``JetPop'' and
PIRSES-GA-2012-31578 ``EuroCal''.
This research was supported in part by NASA grant NNX11A043G and NSF grant AST-1109911, and by the
Polish National Science Centre, grant numbers 2011/01/B/ST9/04618 and  2017/25/B/ST9/02805.
K.\,T. and G.\,P. acknowledge support by the European Commission Seventh Framework Programme (FP7) through
the Marie Curie Career Integration Grant PCIG-GA-2011-293531 ``SFOnset''.
A.N.R., G.P. and A.C.S.R. acknowledge support from the National Science Foundation, under grant number AST-1611547. K.\, T. and D.\,B. acknowledge support from the European Research Counsil under the European Union's Horizon 2020 research and innovation programme, grant agreement No 771282.
M.\,B. acknowledges support from the International Fulbright Science and Technology Award. I.M. was funded by the International Max Planck Research School (IMPRS) for Astronomy and Astrophysics at the Universities of Bonn and Cologne.
T.\,H. was supported by the Academy of Finland project number 317383.
This research made use of Astropy, \url{http://www.astropy.org}, a community-developed core Python package for Astronomy.

\bibliographystyle{mnras}
\bibliography{bibliography_manual}


\label{lastpage}

\end{document}